\documentclass[notitlepage,aps,prl,reprint,twocolumn,longbibliography,superscriptaddress]{revtex4-1}
\usepackage{graphicx}
\usepackage{amsmath}
\usepackage{amssymb}
\usepackage{comment}
\usepackage[colorlinks, allcolors=blue]{hyperref}
\usepackage[all]{hypcap}
\usepackage[mathlines]{lineno}
\usepackage{braket}
\usepackage{wrapfig}
\usepackage{lipsum}
\usepackage{ulem}

\newcommand{\squig}{{\raise.17ex\hbox{$\scriptstyle\sim$}}}

\newcommand{\pref}[2]{\hyperref[#1]{\ref{#1}(#2)}}
\newcommand{\preff}[2]{\hyperref[#1]{\ref{#1}#2}}
\newcommand{\eqpref}[1]{\hyperref[#1]{(\ref{#1})}}

\begin{document}
\title{Collective magnetism of atomic momentum states}
\author{Garrett R. Williams}
\affiliation{Department of Physics, University of Illinois at Urbana-Champaign, Urbana, IL 61801-3080, USA}
\author{Rishi P. Lohar}
\affiliation{Department of Physics, University of Illinois at Urbana-Champaign, Urbana, IL 61801-3080, USA}
\author{Tao Chen}
\affiliation{Department of Physics, University of Illinois at Urbana-Champaign, Urbana, IL 61801-3080, USA}
\author{Brian L. DeMarco}
\email{bdemarco@illinois.edu}
\affiliation{Department of Physics, University of Illinois at Urbana-Champaign, Urbana, IL 61801-3080, USA}
\author{Bryce Gadway}
\email{bgadway@illinois.edu}
\affiliation{Department of Physics, University of Illinois at Urbana-Champaign, Urbana, IL 61801-3080, USA}
\affiliation{Department of Physics, The Pennsylvania State University, University Park, Pennsylvania 16802, USA}
\date{\today}

\begin{abstract}
\noindent
Organization and ordering from interactions in many-body systems underlies our understanding of phases of classical and quantum matter. Magnetism has played a particularly foundational role in the study of many-body phases.
Here, we explore the collective magnetism that emerges from two laser-coupled momentum modes of a scalar bosonic quantum gas. We employ adiabatic state preparation and explore the collective magnetization response to an applied bias potential, finding that the relative increase of interactions leads to an enhanced and muted response for the ground state and excited state, respectively. We further find evidence for significant $Z_2$ symmetry breaking of the sample magnetization for the ground state, consistent with the expected beyond-mean-field behavior. These results suggest that the nonlinear interactions of scalar Bose condensates could provide a simple, direct path towards the squeezing of momentum states for quantum sensing.
\end{abstract}
\maketitle

The coherent control of atomic momentum states has played an important role in the development of ultracold gases, with atom interferometry providing a foundational and enduring application of cold atoms~\cite{Chebotayev:85,CLAUSER1988262,Bragg-Martin,BORDE198910,Atom-Inter-Raman,Cronin-review}. Over the past few decades, the momentum degree of freedom of ultracold gases has further provided a playground to explore momentum-space analogs of myriad transport phenomena~\cite{Moore,steck-dyn,Garreau-Anderson1,Garreau-Anderson-2,Gadway-KR,SeeToh2022,Cao2022,Da-Wei-srl,Gadway-KSPACE,Meier-AtomOptics,An-FluxLadder,Meier-TAI,Liang-NHSE,Liang2024,Agrawal-2d}.
Just as in real-space studies, nonlinearities due to atomic scattering naturally influence the dynamics in these momentum-space systems~\cite{Rolston2002,Ozeri-RMP}. Such interactions have been observed to enrich the dynamics in momentum-space double wells~\cite{An-Inter,GuanSwallow,Engels-Jos} and lattices~\cite{Xie-TopQuantWalks,Alex-nonlinear,An-GAA,Wang-AA-selfdual,Jia-frustr}, albeit at a mean-field level akin to a classical nonlinearity.
While strong interactions between momentum states leading to quantum correlations have recently been realized~\cite{Esslinger-dynamical,Thompson-Luo-2024} for atoms in optical cavities~\cite{Tino-cavity}, it otherwise remains an open challenge to explore beyond-mean-field effects in momentum-state lattices.

\begin{figure}[t]	\includegraphics[width=1.00\columnwidth]{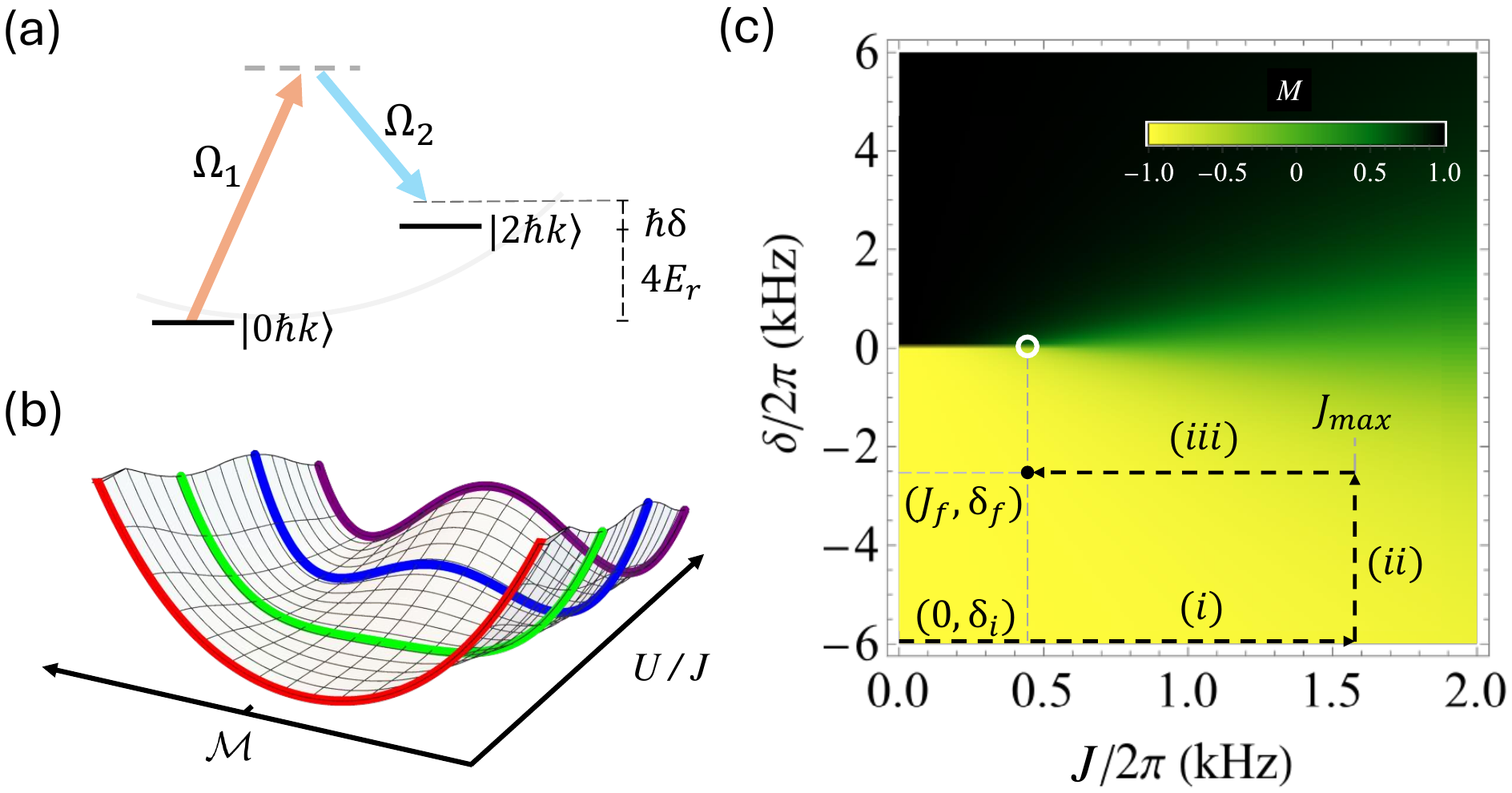}
	\caption{\label{FIG:fig1}
	\textbf{Collective magnetism and symmetry breaking in a momentum-space double well.}
	\textbf{(a)}~A pair of Bragg lasers with Rabi strengths $\Omega_1$ and $\Omega_2$ couple discrete atomic momentum states, $\ket{0 \hbar k}$ and $\ket{2 \hbar k}$, creating an effective double well in momentum space with coherent coupling rate $J \propto \Omega_1 \Omega_2$ and an inter-well bias $\delta$.
    \textbf{(b)}~Free energy landscape at resonance ($\delta = 0$) of the system's ground state (GS) as a function of the magnetization $\mathcal{M}_\textrm{GS} = P_{2} - P_{0}$ and the interaction-to-coupling ratio $U/J$ (with $U$ the scale of collective momentum mode-dependent interactions). An increase of $U/J$ leads to a transition from a unique GS to a bifurcated landscape with two degenerate minima.
    \textbf{(c)}~GS magnetization
    as a function of the coupling $J$ and bias $\delta$, shown for $U/2\pi=$~0.86~kHz.
    The dashed lines indicate the paths of dynamical parameter evolution [(i), then (ii), then (iii)] used to prepare dressed eigenstates in the experiment.
    The white circle ($\delta = 0$, $J = U/2$) indicates the point below which the GS free energy landscape bifurcates
    and $\partial\mathcal{M}/\partial \delta$ diverges.
	}
\end{figure}

Here, we explore the physics of interaction-driven symmetry breaking with a bulk gas of atoms that effectively reside in a momentum-space double well. Considering the momentum double-well as an effective pseudospin, we use adiabatic state preparation to study the atoms' equilibrium collective magnetization in the presence of an external field (bias potential). We observe an enhanced spin susceptibility of the ground state due to collective interactions, contrasted with a reduced susceptibility when preparing the excited state of the system. Finally, we explore higher moments of the ground state magnetization at zero bias field, observing bifurcated collective spin distributions below a critical transverse field. This observation is consistent with an expected spontaneous $Z_2$ symmetry breaking, and suggests that ultracold mixtures of momentum states provide a direct, generic path towards the squeezing of atomic momentum states.

Figure~\ref{FIG:fig1} summarizes the central aspects of our study. As depicted in Fig.~\pref{FIG:fig1}{a}, we couple
atomic momentum modes $\ket{0 \hbar k}$ and  $\ket{2 \hbar k}$
to form an effective double-well system, with the tunneling rate $J \propto \Omega_1 \Omega_2$ and bias $\delta$ set by the Rabi rates and relative detuning ($\delta + 4 E_r/\hbar$) of two counter-propagating Bragg laser fields having wavevectors of magnitude $k$.
While there is a unique single-atom ground state, strong interactions modify this picture via an effective collective ferromagnetism of the two-level system.
As shown in Fig.~\pref{FIG:fig1}{b}, for sufficiently strong interactions ($U/J > 2$, with $\hbar U$ the condensate mean-field energy when a single momentum mode is occupied), the ground state free energy bifurcates into two degenerate minima, relating to a $Z_2$ symmetry breaking.
Figure~\pref{FIG:fig1}{c} considers the effective ground state magnetization
as a function of the Bragg detuning $\delta$ and the coupling rate $J$
(for a fixed interaction strength $U/2\pi = 860$~Hz). As depicted, the magnetization $\mathcal{M}$ becomes more sensitive to the detuning as $J$ decreases ($U/J$ increases), with the magnetic susceptibility diverging when $U/J > 2$.
In the experiments that follow, we probe the equilibrium properties of this system by performing adiabatic sweeps of the Bragg detuning ($\delta$) and coupling strength ($J$), as depicted by the dashed-line paths in Fig.~\pref{FIG:fig1}{c}. 

To capture the critical behavior of Bragg-driven condensate atoms, we map the two-mode problem of $N$ interacting bosons to that of a single total spin $S = N/2$ particle~\cite{RAGHAVAN2001149}. The Hamiltonian can be expressed (up to constant terms) in terms of spin-$N/2$ operators as
\begin{equation}\label{two_mode_ham}
   H_{\mathrm{eff}}/\hbar = \delta \hat{S}_z  + 2J \hat{S}_x -u \hat{S}_z^2.
\end{equation}
Here, $u = U/N$, where $\hbar U = g \rho = (4\pi \hbar^2 a_s/ m)\rho$ is the mean-field energy when all atoms reside in a single momentum state (with $\rho$ the atomic density, $m$ the particle mass, and $a_s \approx 5.3$~nm~\cite{scatt} the $s$-wave scattering length for $^{87}$Rb)~\cite{An-Inter,Alex-nonlinear,SuppMats3}.
The linear terms $\propto$ $\hat{S}_z$ and $\hat{S}_x$
reflect the
Bragg detuning and
Bragg Rabi coupling, respectively.
The one-axis twisting~\cite{Kitagawa-squeeze} term $\propto$ $\hat{S}_z^2$ stems from the mode-dependent scattering of identical bosons in distinguishable momentum modes~\cite{Ozeri-RMP}. For positive $a_s$, atoms in distinguishable modes experience an added repulsive exchange interaction, resulting in an effective local attraction between atoms in momentum space.

Previous studies of the pseudo-magnetism of scalar matter waves have been performed in driven optical lattices~\cite{Madison-shake,Gemelke-shake}, where an
Ising ferromagnetism arises for Floquet-Bloch states~\cite{Parker2013,Clark-scaling} (cf. Ref.~\cite{Campbell-Hysteresis} for pseudomagnetism of angular momentum states). Transverse field terms ($\propto \hat{S}_x$) that establish coherence between degenerate spin configurations do not arise naturally in the driven-lattice setting.
While the introduction of such terms 
via the Bragg-coupling of Raman-dressed states
has recently been proposed~\cite{Zhang-Jos} and realized~\cite{Engels-Jos}, 
the use of Raman-dressed states in this approach leads to a reduced effective interaction strength and a sensitivity to magnetic fields.
Here, we consider how simple Bragg-coupled scalar matter waves allow for the exploration of collective magnetism with a tunable transverse field.

\begin{figure}[b]
\includegraphics[width=1.00\columnwidth]{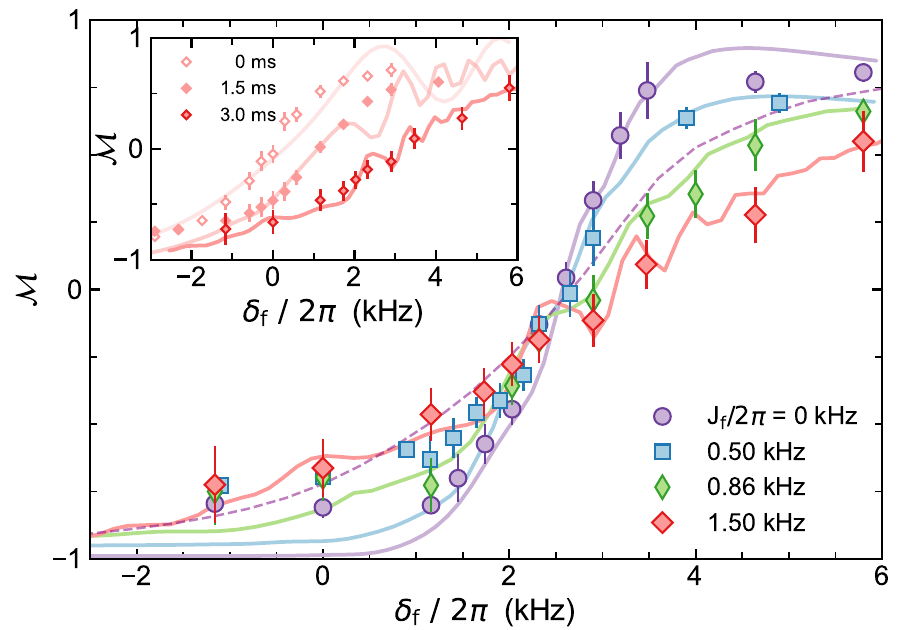}
	\centering
	\caption{\label{FIG:fig2}
		\textbf{Magnetization of the Bragg-dressed ground state (GS).}
        Data points are the measured magnetization of the ramp-prepared GS (determined by fits to the shot-averaged integrated optical density profiles~\cite{SuppMats3})
        for final values of the Bragg coupling $J_f$ as defined in the legend.
        Solid curves are real-space 3D-GPE simulations that incorporate atomic interactions~\cite{Chen2021-gpe,SuppMats3}, while the dashed-line curve relates to the non-interacting result for $J_f = 0$.
        The inset depicts how
        the magnetization response of the dressed GS evolves when simply holding at $J_f / 2\pi = 1.57(4)$~kHz for times 0, 1.5, and 3~ms after the initial 2~ms-long preparation, due to effects of the trapping potential.
        Error bars reflect propagation of standard errors from fits to $\sim$20 momentum profiles.
        }
\end{figure}

Our experiments begin by preparing
samples of $N\approx 8(2) \times 10^4$ Bose-condensed $^{87}$Rb atoms in an optical dipole trap (wavelength 1070~nm) having harmonic trapping frequencies $\omega_{\{x,y,z\}}/2\pi \approx \{130,85,30\}$~Hz.
The nearly-pure Bose-Einstein condensates (BEC) have a calibrated average mean-field interaction energy $\hbar\bar{U} = g\bar{\rho} \approx h \times 900(100)$~Hz~\cite{SuppMats3}, with a corresponding peak value $U_p = (7/4) \bar{U}$~\cite{Stenger-Bragg}. As summarized earlier and depicted in Fig.~\pref{FIG:fig1}{a}, counter-propagating Bragg lasers with a wavelength of $\lambda = 1064$~nm
are used to engineer our two-level system of coupled momentum modes.
When the two laser fields have a frequency mismatch of $4E_{r} / h \approx 8.11$~kHz (with $E_{r} = h^2/2m\lambda^2$ the photon recoil energy), the resultant moving optical potential drives Bragg transitions~\cite{Kozuma-bragg} that preserve the atomic internal state but resonantly couple linear momentum states with momenta $0\hbar k$ and $2\hbar k$ ($k = 2\pi/\lambda$) along the laser propagation direction. By operating in the vicinity of this resonance, with the Bragg detuning from resonance, $\delta/2\pi$, and the two-photon Bragg coupling rate, $J/2\pi$, kept small in comparison to Doppler shift from other allowed resonances (spaced by $8E_r/h = 16.22$~kHz), we can consider the atoms to reside in an effective two-level system.

To describe the collective magnetism of this two-level system, we define the sample magnetization as $\mathcal{M} = -2 \langle S_z \rangle / N \equiv P_{2} - P_{0}$, where $P_{n}$ is the normalized population in the mode with momentum $n \hbar k$. To experimentally probe the collective magnetism of the ground state of this system, as well as the most-excited state, we prepare optically-dressed eigenstates in the limit of large $J$ and then slowly increase the ratio $U/J$ to explore the influence of atomic interactions.
While our use of Bragg-coupled momentum states comes with some attractive features,
it also presents challenges to adiabatic state preparation. Our condensate atoms have a finite spatial extent, and the momentum modes $0 \hbar k$ and $2 \hbar k$ will lose overlap if left to freely evolve. Bragg-dressed states (for large $J$) will be robust to this spatial separation, however. Our harmonic trapping potential
does still complicate the evolution of the Bragg-dressed states. 
While the trap forces will be small for timescales much less than a quarter trap period ($T/4 \approx 8.3$~ms), the dressed states will still experience a Doppler shift of their two-photon resonance as they are slowed by the trap.

To ameliorate these influences, we perform an abbreviated state preparation procedure as follows [indicated by paths (i-iii) in Fig.~\pref{FIG:fig1}{c}]. First, over 2~ms, we prepare simple (coherent state) Bragg-dressed eigenstates in the large-field limit, keeping the rms Bragg ``field'' $\sqrt{\delta^2 + 4J^2}$ large compared to the collective nonlinearity $\bar{U}$. We first maintain a large initial Bragg detuning $\delta_i/2\pi = -5.8$~kHz while smoothly ramping up $J$ from 0 to $J_{max}/2\pi = 1.57(4)$~kHz
over 1~ms.
Then, we linearly ramp $\delta$ over 1~ms to a tunable value $\delta_f$ in the vicinity of the Bragg resonance.
We conclude with a 3~ms-long ramp-down of $J$ from $J_{max}$ to $J_f$ to probe the atoms' collective magnetic response under an increasing $U/J$.

Figure~\ref{FIG:fig2} shows the resulting magnetization vs. bias ($\mathcal{M}$ vs. $\delta_f$) response of the ground state for several $J_f$ values.
For large $J_f$, we observe a broad sigmoidal response as one would expect for a non-interacting paramagnet, with the magnetization taking values close to $\pm 1$ when the bias dominates over the transverse field ($|\delta_f / J_f| \gg 1$).
With decreasing $J_f$ (increasing $U/J$), we observe a systematic sharpening of the magnetization response curve.
To assess the impact of interactions on this increased sensitivity to the applied bias, we compare to theory simulations based on a three-dimensional Gross-Pitaevskii equation (3D-GPE)~\cite{Chen2021-gpe}, which also naturally account for effects of the trapping potential, including the inhomogeneous atomic density~\cite{SuppMats3}.
We find good agreement between the data and theory
when we account for atomic interactions (solid curves, based on a global agreement between the data and theory for an average mean-field interaction $\bar{U}/2\pi = 860$~Hz, consistent with independent calibration measurements~\cite{SuppMats3}).
We find a much poorer agreement if we assume a non-interacting gas (dashed line for the $J_f = 0$ ramp).

We note one peculiar feature of the data in Fig.~\ref{FIG:fig2} that is captured by our 3D-GPE simulations - a shift of the nominal $\mathcal{M} = 0$ point away from the expected Bragg resonance condition $\delta_f = 0$ by roughly 2-3~kHz. Because our sweep from $\delta_i$ to $\delta_f$ occurs when $J$ is much larger than $\bar{U}$, this shift does not stem from interactions (i.e., a mean-field shift). Rather, the shift appears because our prepared ground dressed state
moves in the trap, gets slowed down by the trap, and experiences an effective Doppler shift of its resonance.
The observed shift is fully captured by the 3D-GPE, which incorporates the 30~Hz trapping frequency along the direction of imparted momentum as well as the full state preparation procedure.
In the inset of Fig.~\ref{FIG:fig2}, we directly confirm the role of the in-trap dynamics, comparing how the $J_f = J_{max}$ (no $J$ rampdown) magnetization curve evolves as we simply hold for 0, 1.5, and 3.0~ms after the first 2~ms of state preparation. The curves shift, in agreement with the simulations, as the dressed GS evolves over the hold time.

\begin{figure}[b]
	\includegraphics[width=0.98\columnwidth]{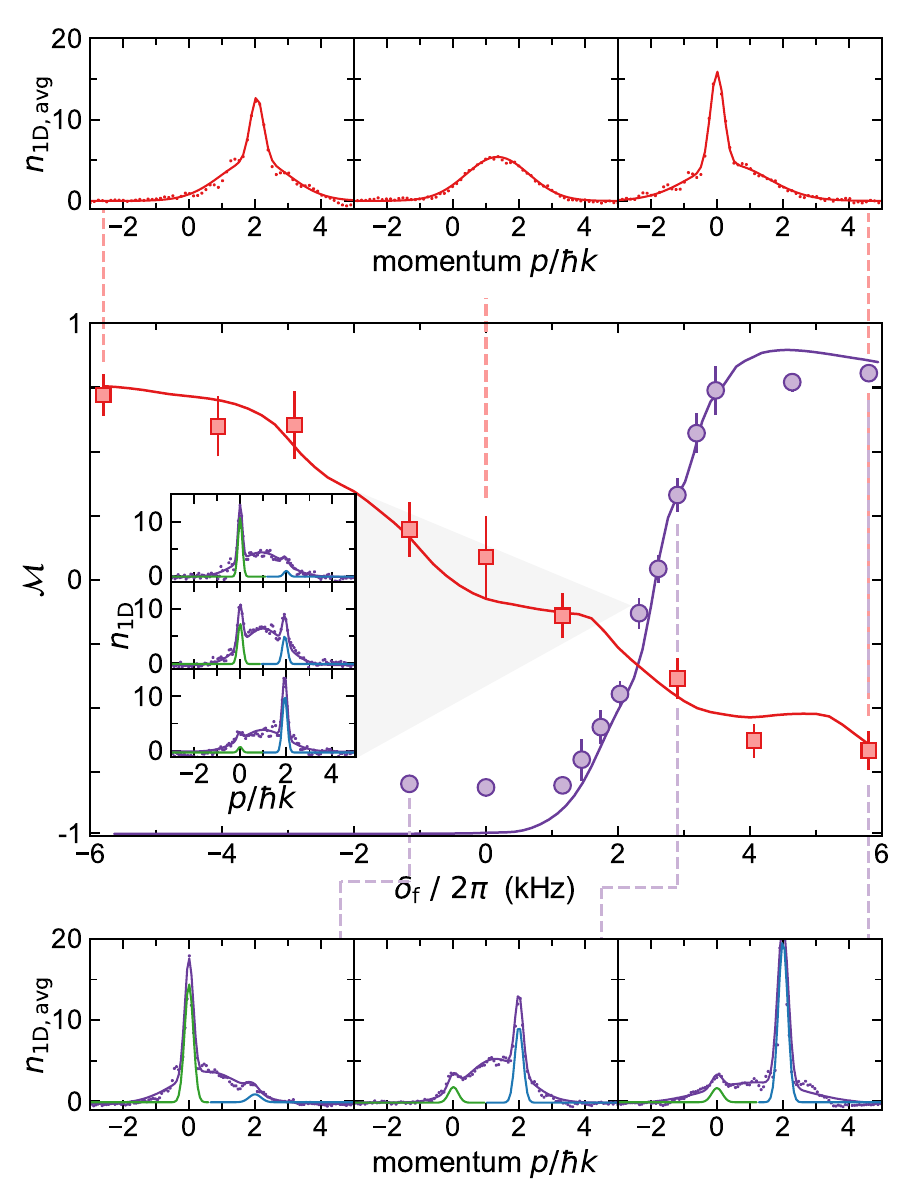}
	\caption{\label{FIG:fig3}
	\textbf{Comparison of ground state (GS) and excited state (ES) magnetization.}
       Data points are the fit-determined mean magnetization of the GS (purple circles, as in Fig.~\ref{FIG:fig2}) and ES (red squares) vs. the final detuning $\delta_f$.
       Both the ES and GS data relate to $J_f = 0$, respectively prepared by $\delta_i/ 2 \pi = \pm 5.8$~kHz.
       At the top and bottom, averaged (over $\sim$20 images, same as the data points) integrated optical density (OD) profiles are shown for the ES and GS, respectively, at indicated $\delta_f$ values.
       The GS
       magnetization is determined~\cite{SuppMats3} by the normalized difference between the integrated weight under the $2 \hbar k$ (blue) and $0 \hbar k$ (green) fit peaks [$\mathcal{M}_{\rm GS} = P_2 - P_0$, where $P_0 + P_2 = 1$].
       For the ES, where only a single, broad peak is observed near resonance, the magnetization is determined~\cite{SuppMats3} by the differential fraction of the overall fit distribution
       having momentum above and below $1 \hbar k$ [$\mathcal{M}_{\rm ES} = P_{>1} - P_{<1}$, where $P_{>1} + P_{<1} = 1$].
       Solid curves are time-dependent 3D-GPE simulations~\cite{Chen2021-gpe,SuppMats3}.
       The inset depicts characteristic single-shot OD profiles for $\delta_f/ 2\pi = 2.32$~kHz, near      
       the trap-shifted resonance.
       Error bars reflect propagation of standard errors from the fits.
}
\end{figure}

\begin{figure*}[htb!]
\begin{center}
\includegraphics[width=2.0\columnwidth]{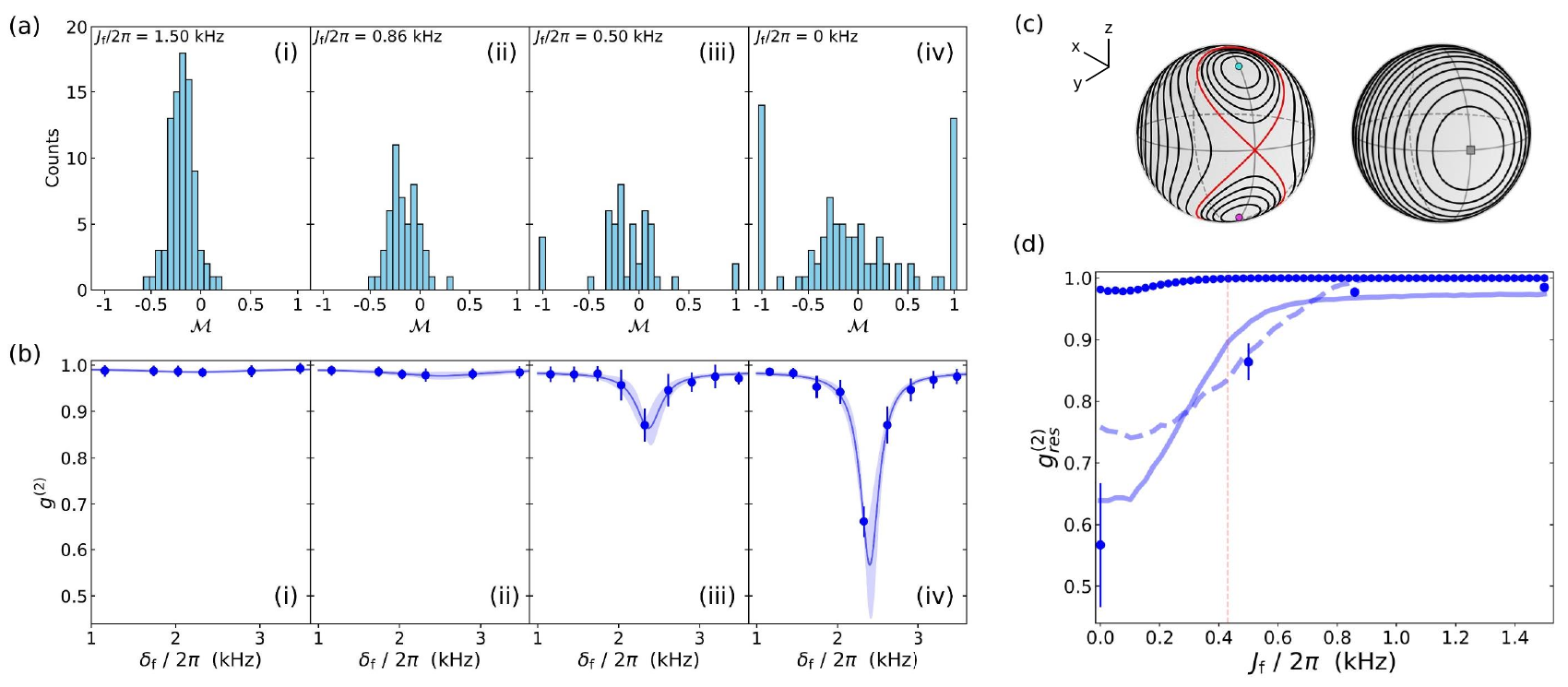}
\caption{\label{FIG:fig4}
		\textbf{Spontaneous symmetry breaking of the Bragg-dressed ground state.}
        \textbf{(a,i-iv)}~Distributions of the single-shot $\mathcal{M}_{\rm GS}$ for $J_f/2\pi$ values of 1.50~kHz,  0.86~kHz, 0.50~kHz, and 0~kHz, respectively.
        \textbf{(b,i-iv)}~Two-mode correlation function $g^{(2)} = \braket{P_{0 \hbar k}P_{2 \hbar k}}/\braket{P_{0 \hbar k}}\braket{P_{2 \hbar k}}$ for the ramp-prepared GS vs. $\delta_f$, based on distributions as in (a). Error bars are based on bootstrapping of the $\mathcal{M}_{\rm GS}$ distributions. 
        \textbf{(c)}~Mean-field phase space contour plot for $J > J_c$ (right) and $J < J_c$ (left). Above $J_c = \bar{U}/2$, there is a unique ground state represented by the grey square at $\mathcal{M} = 0$.
        Below $J_c$, the GS phase space is divided by a separatrix (red line),     bifurcating into two wells at $\mathcal{M}_{\pm} = \pm \sqrt{1 - J^2/J_c^2}$
        denoted by cyan and magenta points.
        \textbf{(d)}~The on-resonance correlation $g^{(2)}_{res}$ vs. $J_f$.
        Error bars are the standard error of Lorentzian fits used to extract $g^{(2)}_{res}$ as shown (with 90\% confidence regions) in (b).
        The solid blue line is based on a two-mode truncated Wigner approximation
        assuming an initial Gaussian phase space distribution
        with standard deviation $\sigma_\mathcal{M}$ = 0.15 and mean interaction strength $\bar{U}$ = 2$\pi \times$ 0.86 kHz~\cite{SuppMats3}. The dotted and dashed lines are obtained using the same method but assuming a coherent state distribution with $\sigma_M = \sqrt{2/N} = 1/200$ and with average mean field energies $\bar{U}$ and $3 \bar{U}/2$ respectively.
        The red vertical line indicates $J_c = \bar{U}/2$. 
}
\end{center}
\end{figure*}

We more directly illustrate the influence of atomic interactions by contrasting the magnetization response of the ground state to that of the \textit{excited} dressed state.
The observable properties of the most excited eigenstate of Eq.~\ref{two_mode_ham} for a given $\{\delta$,$u\}$ are equivalent to those of the ground state for the negated parameters, $\{-\delta$,$-u\}$. 
We prepare the dressed excited state (ES) via the same ramping procedure as for the ground state (GS), but starting instead with an initial Bragg detuning of $\delta_i = +5.8$~kHz.
Figure~\ref{FIG:fig3} contrasts the response of the ES (red) to that of the GS (purple, same data and curve as in Fig.~\ref{FIG:fig2}), both for a full ramp down to $J_f = 0$.
For the ES response, we observe the expected change in sign of the slope with respect to $\delta$, but also find that the magnitude of the susceptibility ($|\partial \mathcal{M}/\partial \delta_f|$) becomes much reduced, reflecting an inhibited alignment with the external bias field due to interactions. This is in direct contrast to the GS's enhanced tendency to align with the external bias field. 
In other words, interactions make the GS more pliable to the external bias field, while the ES response is stiffened. These contrasting responses reflect the evolution of the normal modes of this two-mode nonlinear system~\cite{Raghavan}: the free energy landscape of the ES becomes stiffer with increasing $U/J$, whereas the GS is softened, becoming an unstable fixed point for $U/J > 2$ (cf. Fig.~\pref{FIG:fig1}{a}).

In addition to fit-determined $\mathcal{M}$ values, we also plot some averaged (over $\sim$20 shots) integrated time-of-flight (TOF) density profiles of the GS (bottom panel) and ES (top panel) for different $\delta_f$ values. We note the qualitative differences between the GS and ES, with the GS always exhibiting narrow discrete momentum peaks on top of a small background (from thermal atoms and $s$-wave scattering) while the ES lacks narrow peaks near resonance. This absence of narrow momentum orders relates to the onset of an instability in the ES, which we observe also in our 3D-GPE simulations~\cite{SuppMats3}. Thus, our determination of $\mathcal{M}$ for the ES, detailed in the Fig.~\ref{FIG:fig3} caption, does not rely on peak fitting as for the GS~\cite{SuppMats3}.

We further plot typical individual TOF profiles for the GS near resonance (central inset to Fig.~\ref{FIG:fig3}). We observe large shot-to-shot fluctuations, with many profiles exhibiting near-full polarization in either the $\ket{0 \hbar k}$ or $\ket{2 \hbar k}$ peak. These fluctuations of the single-shot GS magnetization are consistent with the picture that the prepared GS resides at an unstable fixed point when $U/J > 2$.
We note that the degree of symmetry breaking we observe for such a short state preparation ramp is in contrast to the partial symmetry breaking and domain formation found in related experiments~\cite{Parker2013,Clark-scaling} on Floquet-engineered Bloch states~\cite{Madison-shake,Gemelke-shake}, suggestive of the
important role played by our coherent transverse Bragg fields.

Figure~\ref{FIG:fig4} analyzes the fluctuations of the GS response in more detail, providing some evidence for a spontaneous $Z_2$ symmetry breaking.
Figure~\pref{FIG:fig4}{a} shows histograms of the sample GS magnetization on resonance (at the trap-shifted resonance of $\delta_f/2\pi = 2.32$~kHz) for several values of $J_f$. As $J_f$ is lowered, the histograms transition from a single $\mathcal{M}$~$\approx$~$0$ peak to bifurcated distributions with peaks near $\mathcal{M}$~$\approx$~$\pm 1$.
From such distributions, we determine the modal correlation $g^{(2)} = \braket{P_{0 \hbar k}P_{2 \hbar k}}/\braket{P_{0 \hbar k}}\braket{P_{2 \hbar k}} = \braket{1-\mathcal{M}^2}/\braket{1-\mathcal{M}}\braket{1+\mathcal{M}}$, where brackets denote an average over the set of realizations. Figure~\pref{FIG:fig4}{b} shows the trend of $g^{(2)}$ as a function of $\delta_f$ and $J_f$, indicating that modal anti-correlations appear near resonance for small $J_f$ values, consistent with the expectation of $Z_2$ symmetry breaking below a critical coupling.
Figure~\pref{FIG:fig4}{c} depicts the phase-space contours of the two-mode mean-field model, indicating the GS transition from a stable (right, $J > J_c$) to an unstable (left, $J < J_c$) fixed point.

Figure~\pref{FIG:fig4}{d} summarizes the onset of GS anti-correlations as the $J_f$ values drop below a critical Bragg coupling strength 
(dashed vertical line indicating
$J_c = \bar{U}/2 = 2\pi \times 430$~Hz), 
plotting the maximal $g^{(2)}$ values (from Lorentzian fits in Fig.~\pref{FIG:fig4}{b}) vs. $J_f$.
While an exact theory comparison is challenged by the multi-mode nature
of our many-particle system, we provide some approximate comparisons that capture beyond-mean-field ($g^{(2)} \neq 1$) effects by incorporating fluctuations in the spirit of the truncated Wigner approximation (TWA).
Specifically, we consider the classical limit of Eq.~\ref{two_mode_ham}, initializing different states (from a coherent state distribution) in the two-mode phase space and evolving according to the experimental 3~ms-long ramp down of $J$ to $J_f$.
We incorporate some static effects of the trap, i.e. the inhomogeneous density, by sampling over a distribution of $U$ values as in a local density approximation.

The dotted curve shows the result for the expected (mean value $\bar{U}/2\pi = 860$~Hz) interaction distribution and for the phase-space distribution relating to a coherent state (CS) of $8\times 10^4$ particles. Only a slight deviation from the uncorrelated value  ($g^{(2)} = 1$) is observed, stemming from the finite duration of the ramp.
We can find better agreement with the data by either assuming an enlarged interaction value (dashed curve, sampling over a distribution with mean value $1.5\bar{U}$) or by assuming the expected interaction strength but sampling from a phase-space distribution significantly larger than that of the expected CS (solid curve)~\cite{SuppMats3}.
These trends could suggest 
a failure of our two-mode theory to describe the data and its analysis, 
an underestimation of $\bar{U}$,
or additional effects related to, e.g., finite temperature, trap dynamics (including the formation of magnetic domains~\cite{Parker2013}), or shot-to-shot parameter variations.
Future work
will seek to sort out these influences,
also incorporating slower, more adiabatic ramping protocols
based on the use of
stationary dressed states (superpositions of $\pm 1 \hbar k$).

These results demonstrate that atomic momentum states provide a setting to explore the physics of collective quantum magnetism, with hints of beyond-mean-field correlations.
Broadly speaking, momentum state mixtures should be analogous to their internal state counterparts~\cite{DSK-Ueda-spinor}, and one may hope that
they will enjoy a similar 
steady progression from mean-field dynamics~\cite{Chapman-Spinor,Zibold,Ferrari-PRA} to studies of quantum correlations~\cite{Mu-sque,Hoang2016,Luo620,Qu-gerb}. This optimistic analogy would suggest that contact interactions in scalar gases could be used to directly engineer many-body entangled states of atomic momentum relevant for quantum-enhanced sensing, without the use of internal spin states~\cite{klempt1,Greve2022}.
Many body momentum-space systems may additionally provide a venue to explore foundational questions regarding the equilibrium behavior of attractively interacting bosonic gases in degenerated landscapes~\cite{Noziéres_1995}. Where attractive real-space interactions lead to instability and collapse, stable effective attraction in momentum space may enable the exploration of analogous problems in momentum space~\cite{Higbie,Tomoki-rashba,Sarang-Rashba,Qi-Rashba}.

\section{Acknowledgements}
We thank Junpeng Hou, Chuanwei Zhang, and Qi Zhou for helpful and stimulating discussions.
This work was supported by the Air Force Office of Scientific Research under Grant No. FA9550-19-1-0272 (G.R.W., R.P.L., B.L.D., and B.G.) and under Grant No.~FA9550-21-1-0246 (T.C. and B.G.).
G.R.W. acknowledges support from both the National Science Foundation through the Graduate Research Fellowship Program (NSF-GRFP) and from the Alfred P. Sloan foundation. R.P.L. acknowledges support by the Department of Defense (DoD)
through the National Defense Science and Engineering Graduate (NDSEG) Fellowship Program.

\bibliographystyle{apsrev4-1}
\bibliography{NewBib}

\newcommand{\hH}{\hat{H}}
\newcommand{\hp}{\hat{\psi}}
\newcommand{\+}{^\dagger}
\newcommand{\rmi}{{\rm i}}
\newcommand{\rmd}{{\rm d}}
\newcommand{\br}{\mathbf{r}}
\newcommand{\vk}{\mathbf{k}}
\newcommand{\vp}{\mathbf{p}}
\newcommand{\vq}{\mathbf{q}}

\clearpage

\end{document}